%% file: fse-vision-2016.tex
\documentclass{sig-alternate-05-2015}
\usepackage[latin9]{inputenc}
\usepackage{url}
\usepackage{color}
\usepackage{soul}
\usepackage{booktabs}
\usepackage{graphicx}
\usepackage{multirow}
\usepackage{citesort}
\usepackage{amssymb}
\usepackage{amsbsy}
\usepackage{citesort}

\global\long\def\ib{\boldsymbol{i}}
\global\long\def\fb{\boldsymbol{f}}

\global\long\def\ob{\boldsymbol{o}}

\begin{document}

\setcopyright{acmcopyright}

\doi{10.475/123_4}

\isbn{123-4567-24-567/08/06}

\conferenceinfo{FSE '16}{November 13--19, 2016, Seattle, WA, USA}
\acmPrice{\$15.00}

\title{DeepSoft: A vision for a deep model of software}

\numberofauthors{4} 
%
\author{
	%
	%
	\alignauthor
	Hoa Khanh Dam\\
	\affaddr{University of Wollongong \\Australia}\\
	\email{hoa@uow.edu.au}
	\alignauthor
	Truyen Tran\\
	\affaddr{Deakin University \\ Australia}\\
	\email{truyen.tran@deakin.edu.au}
	\alignauthor
	John Grundy\\
	\affaddr{Deakin University \\ Australia}\\
	\email{j.grundy@deakin.edu.au}
\and
	\alignauthor
	Aditya Ghose\\
	\affaddr{University of Wollongong \\ Australia}\\
	\email{aditya@uow.edu.au}
}

\maketitle
\begin{abstract}
Although software analytics has experienced rapid growth as a research area, it has not yet reached its full potential for wide industrial adoption. Most of the existing work in software analytics still relies heavily on costly manual feature engineering processes, and they mainly address the traditional classification problems, as opposed to predicting future events. We present a vision for \emph{DeepSoft}, an \emph{end-to-end} generic framework for modeling software and its development process to predict future risks and recommend interventions. DeepSoft, partly inspired by human memory, is built upon the powerful deep learning-based Long Short Term Memory architecture that is capable of learning long-term temporal dependencies that occur in software evolution. Such deep learned patterns of software can be used to address a range of challenging problems such as code and task recommendation and prediction. DeepSoft provides a new approach for research into modeling of source code, risk prediction and mitigation, developer modeling, and automatically generating code patches from bug reports.

 \end{abstract}

%
%

%
%



\section{Introduction}

Software analytics has emerged as one of the fastest growing research areas in software engineering in the recent years. This emergence coincides with the constantly increasing number of software products being built -- ``software is eating the world'' \cite{Andreessen2011}, and the huge amount of data generated from the development, maintenance and usage of software. Software analytics allows us to obtain insights from this data and build classification models (e.g. classifying if source code is defective or non-defective \cite{Menzies:2010}) or risk prediction models (e.g. predicting if a software task will be delayed \cite{Choetkiertikul2015}). Using current machine learning techniques, existing work in software analytics builds models for these problems by \emph{manually} designing and extracting features representing parts of a software system or related development process (e.g. a source code file or an bug report) and using them as predictors. For example, the performance of most of existing defect prediction models heavily relies on the careful handcrafting of good features (e.g. code complexity and code churn metrics) which can discriminate between defective source files and non-defective ones \cite{Menzies:2010}.

Coming up with good features is difficult, time-consuming, and requires domain-specific knowledge, and hence poses a major challenge. In many situations, handcrafted features normally do not generalize well: features that work well in a certain software project may not perform well in other projects \cite{Zimmermann:2009:CDP}. A one-size-fits-all approach is therefore inadequate, necessitating bespoke solutions (in a manner akin to  \emph{personalized medicine} \cite{Schork2015}). Manually designing features for each single software system is however expensive and is thus not a sustainable nor a practical solution. In addition, handcrafted features often rely on expert knowledge, which is sometimes based on outdated experience and an underlying bias, thus impeding the discovery of new, useful features. Hence, we believe that the wide adoption of software analytics in industry crucially depends on the ability to\emph{ automatically} derive (learn) features from raw software engineering data.




In addition, most existing work in software analytics focuses on employing the traditional notion of classification (e.g. classifying defect files). While much work has been done on recommender systems, such as for APIs, this has mostly relied on manual feature identification and using classification approaches. Limited work has been done to address a significantly more  difficult problem: \emph{forecasting future risks or events} (e.g. a delayed release or important functionalities omitted from a release) in software and \emph{recommending appropriate interventions,  code patches or other ``fixes''}. Predicting future risks is highly challenging due to the inherent uncertainty, temporal dependencies, and especially the dynamic nature of software. This is exactly the area where software analytics can contribute most, by learning from potentially large datasets and automatically forming a \emph{deep understanding} of the software, the process of building and maintaining it, and its stakeholders. Such an understanding along with relevant memory of past experience will facilitate automated support for risk prediction and interventions.

\begin{figure*}[ht]
	\centering
	\includegraphics[width=0.9\linewidth]{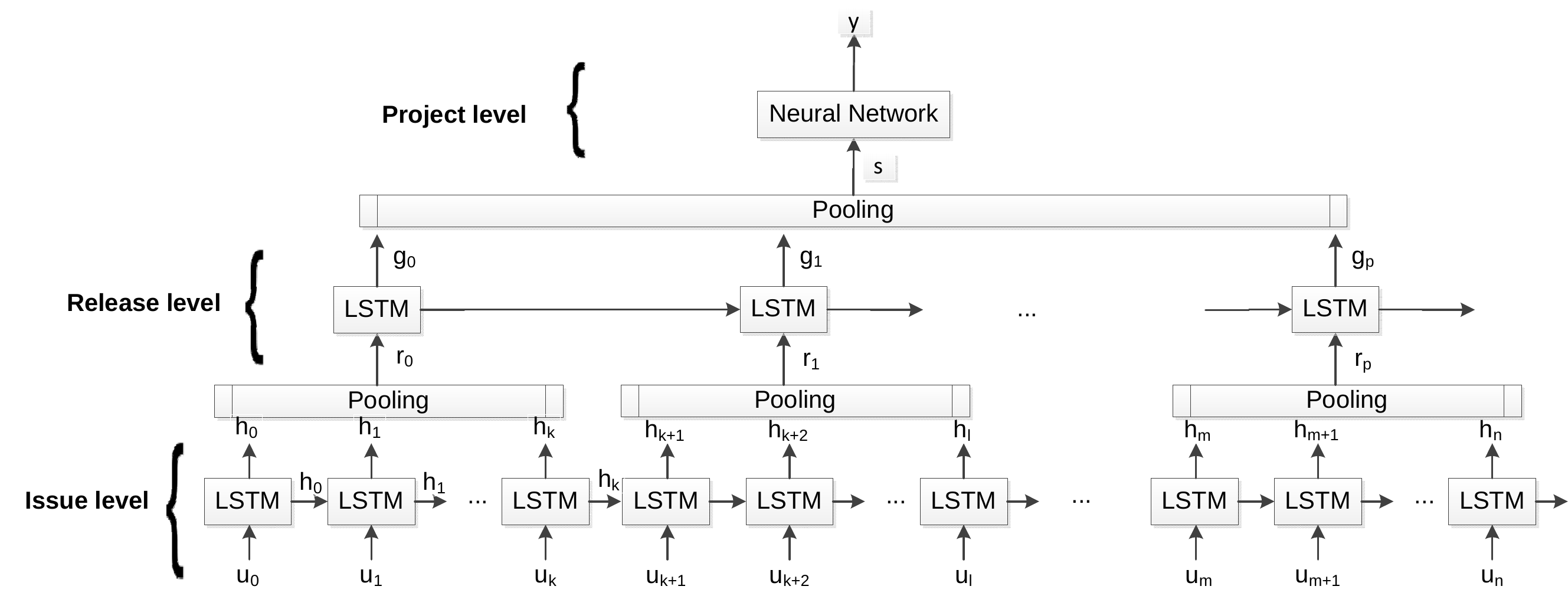}
	\caption{DeepSoft architecture}
	\label{fig:DeepSoft}
\end{figure*}

\section{Challenges and Solution}

Today's software products undergo rapid cycles of development, testing, and release. Development includes implementing new functionalities and fixing bugs, all commonly referred to as resolving \emph{issues}. The resolution of an issue may involve changes to the code in the form of code patches. A release signifies a milestone where a number of issues are resolved. This process is similar to the healthcare process where illnesses are diagnosed and treated, and interventions are put into place to counter future medical risks \cite{DeepCare2016}.

There are four major challenges in providing effective model of the software process: (i) \emph{Handling long-term dependencies in software evolution}: future issues and resolutions may critically depend on historical issues and resolutions. For example, the implementation of a functionality may constrain how other functionalities are implemented in the future. (ii) \emph{Representation of issues and their resolutions}: the challenge here is how features representing the semantics of an issue's description, its diagnoses (e.g. the comments and discussions), and its resolution (e.g. code patch) can be automatically learned from raw data. (iii) \emph{Episodic and irregular timing}: software is iteratively developed and delivered in releases, each of which can be seen as an episode. Time between releases and between the resolution of issues are largely random. Existing learning systems used in software analytics failed to address the episodic and irregular timing of events of interests in software development. (iv) \emph{Modeling confounding interactions between the progression of issues and resolutions}: the phenomenon of bug-introducing changes, where a change to a software system (either to add a new functionality, to restructure the code, or to fix an existing bug) inadvertently injects new bugs, is an example of confounding interactions.  

In this vision paper, we introduce DeepSoft, a generic, dynamic deep learning framework that addresses the above challenges. DeepSoft is developed based on Long Short-Term Memory  \cite{hochreiter1997long}, a recurrent neural network equipped with memory cells to store experiences. DeepSoft is an \emph{end-to-end} prediction model that does not require manual feature engineering. DeepSoft is capable of reading historical software data (e.g. issue reports and source code), memorizing a long history of past experience, inferring the current ``health'' state of a software, predicting future risks, and finally recommending actionable interventions. In the remainder of the paper, we will describe the architecture of DeepSoft and outline a research agenda based on a number of applications of DeepSoft to various software engineering problems.

\input{approach}

\input{applications}

\small
\bibliographystyle{abbrv}

\end{document}

%% file: approach.tex
\section{DeepSoft}

Software is similar to an evolving organism: what will happen next to a software system depends heavily on what has previously been done to it. DeepSoft leverages a deep recurrent neural network (RNN) to model this temporal evolution. Recurrent networks can be seen as multiple copies of the same network, each passing information to a successor and thus allowing information to persist. DeepSoft is built upon Long Short-Term Memory (LSTM) \cite{hochreiter1997long}, a special kind of RNN that is capable of learning long-term dependencies, i.e. remembering information for long periods of time. LSTMs have demonstrated ground-breaking performance in many applications such as machine translation, 
video analysis, 
and speed recognition. 

We focus here on two significant events\footnote{DeepSoft can however be easily extended to model other temporal events.} in the life of a software application: an issue being resolved (which may result in code patches) and a version being released. DeepSoft has several layers (see Figure \ref{fig:DeepSoft}) which model the progression of a software at three levels: issue, release and project. The bottom layer consists of a chain of repeating modules of LSTM units, each of which reads an input $u_t$, representing an issue being resolved at time $t$, and the output $h_{t-1}$ from the previous unit, to compute the output $h_t$. Thus $h_t$ summarizes information from all previous inputs $u_0, u_1, ..., u_{t-1}$. Note that resolving issues can be done interleavedly, and the issues are ordered with respect to their resolved time.

The input $u_t$ represents both the diagnosis of an issue (denoting as vector $x_t$), its resolution $p_t$, and the elapsed time $\Delta_t$ between this issue and the previous one, i.e.  $u_t = [x_t, p_t, \Delta_t]$. The diagnosis of an issue is typically in the form of natural language text capturing its description, the discussion around it (e.g. comments), and optionally some attributes (e.g. type, priority, etc.). State-of-the-art NLP techniques such as word2vec \cite{MikolovSCCD13} and paragraph2vec \cite{Le2014} can be used to automatically convert those texts into a vector which represents the actual semantic of the text. Issue resolutions which result in code patches can also be represented using those NLP techniques since we can view the code as a collection of statements in a language. In the next section, we will discuss in more details how LSTM can also be leveraged to build a deep model for source code.

\begin{figure}[ht]
	\centering
	\includegraphics[width=0.5\linewidth]{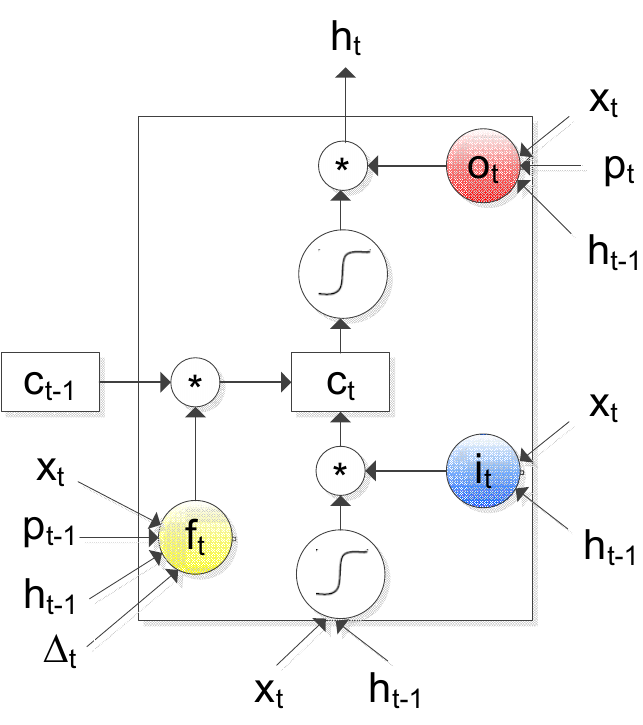}
	\caption{The internal structure of an LSTM for processing a sequence of issues}
	\label{fig:LSTMunit}
\end{figure}

The most powerful feature of an LSTM unit is the \emph{memory cell}  ($c_t$ in Figure \ref{fig:LSTMunit}) that stores accumulated memory of past experience. The amount of information flowing through the memory cell is controlled by three gates (an input gate, a forget gate, and an output gate), each of which returns a value between 0 (i.e. complete blockage) and 1 (full passing through). All these gates are \emph{learnable} (i.e. they can be trained with historical data) in which they are trained with historical data to maximize the predictive performance of the whole model.

We now walk through how we envision an LSTM unit would work in DeepSoft. First, an LSTM unit decides how much information from past experience (i.e. $c_{t-1}$) should be removed from the memory cell. This is controlled by the forget gate $\fb_{t}$, which looks at the the previous output state $h_{t-1}$, what has been done to resolve the previous issue ($p_{t-1}$), the elapsed time $\Delta_t$, and the current issue $x_t$, and outputs a number between 0 and 1. A value of 1 indicates that all the past memory is preserved, while a value of 0 means ``completely forget everything''. For example, if the resolution for the previous issue is marking the issue as ``invalid'', then we may not need to remember this since no changes were made to the software. On the other hand, previous resolutions resulted in code patches should be remembered since those code changes may inject bugs which will later lead to new issues. In DeepSoft, the forget gate is used to model the long-term impacts of issue resolutions. For example, refactoring the code for resolving a technical debt may have long-term benefits (e.g. improving the maintainability of the software) and thus should be remembered. In DeepSoft, the elapsed time is also a factor moderating the forget gate since for example recent changes made to a software may be more relevant than changes made a few years ago. For irregular timing, the forget gate is extended to be a function of irregular time gap between consecutive time steps. A range of forgetting mechanisms can be implemented here. For example, the simple monotonic decay approach mimics the natural forgetting when learning a new concept in human, while the time-parametrization mechanism accounts for more complex dynamics of different issues over time. The resulting model is sparse in time and efficient to compute since only observed records are incorporated, regardless of the irregular time spacing.



The next step is updating the memory with new information obtained from the current issue $x_t$. In a similar manner to the forget gate, the input gate $\ib_{t}$ is used to control which information about the current issue will be stored in the memory. The gate can also modified to reflect the priority level of an issue. For example, the input gate may give a higher value for a major bug issue than for a technical debt issue. Finally, information stored in the memory cell will be used to produce an output $h_t$. The output gate $\ob_t$ looks at the current issue $x_t$ and its resolution $p_t$, the previous hidden state $h_{t-1}$, and determine which parts of the memory should be output. For example, if some of the work done for the current issue is needed for resolving the next issue then that relevant part of the work should be output.

A release requires the resolution of a number of issues. Once the issue dynamics have been modeled using LSTM (refer to the issue level in Figure \ref{fig:DeepSoft}), the next step is aggregating the issue states to model the release dynamics. The aggregation operation is known as pooling. The simplest method is mean-pooling where the vector representing a release (i.e. $r_i$ in Figure \ref{fig:DeepSoft}) is the sum of the output vectors of all the issues in the release divided by the number of issues. More complex pooling methods can be used to reflect the attention to recency such as giving more weight to recent issues than old issues. The sequence of releases $r_0, r_1, ..., r_p$ are also input to another layer of LSTMs which generates a corresponding sequence of states $g_0, g_1, ..., g_p$. These states are aggregated into a single vector $s$ (also using pooling mechanisms) which represents the state of the whole project. Vector $s$ is fed into a neural network which is trained to predict future events.

%% file: applications.tex
\section{Applications}

DeepSoft is a compositional architecture in which an output from a module (e.g. an LSTM unit) can be used as input for the next module (e.g. another LSTM unit). In addition, DeepSoft provides a holistic vector representation of software and its entire evolution. At the issue level (see Figure \ref{fig:DeepSoft}), vector $h_t$ captures not just only information about the current issue but also the knowledge of the previous issues and what have been done to resolve them. Vector $g_j$ represents information about all the past releases, while vector $s$ represent the state of the entire project.  Those vector representations are powerful since they are mathematically and computationally convenient for machine learning algorithms to process in building predictive models at different levels: issues, releases, and projects. Although DeepSoft addresses only three constructs, these are general enough to accommodate most of the types of analysis traditionally considered within the ambit of software analytics. We therefore envision many applications of the DeepSoft to software engineering problems ranging from requirements to maintenance.


\subsection{Source code modeling}


DeepSoft requires a vector representation of an issue and its resolution. Since the description of an issue is in natural language, state-of-the-art deep learning-based NLP techniques \cite{manning2016computational} such as \emph{word2vec}, \emph{paragraph2vec} or Convolutional Neural Networks (CNNs; used in the ground-breaking Facebook's DeepText engine) can be leveraged to \emph{automatically} embed an issue description into a vector representation. A meaningful vector representation for issue resolutions, which are usually in the form of code patches, requires modeling of source code. Code elements such as tokens, methods, and classes are embedded into vectors.  A code token can be treated in one of the following ways: (a) as a sequence of characters, (b) as an atomic unit, or (c) as a member of a more abstract unit, such as elements in an Abstract Syntax Trees (AST). For the first two cases, code is treated in a sequential manner as natural languages, and existing deep learning-based NLP techniques such as \emph{word2vec}, \emph{paragraph2vec} or CNN can be applied here (see \cite{Nguyen:2016} for an example). Since LSTM is specifically designed for modeling sequences (as seen in DeepSoft), it can also be used to model code. For case (c), a piece of code could be a graph, which can be represented either using graph-based LSTM or CNN (e.g., see \cite{mou2016convolutional} for CNN on top of AST). Vector-based embedding enables simple ways for complex inference such as similarity search, grouping, analogy reasoning and token prediction. A major challenge remains to capture the complex interaction between various code composites e.g., methods, classes, modules, components, inheritances and call graphs.


\subsection{Resolution recommendation}



Issue resolutions can range from marking an issue as ``invalid'' or ``duplicate'' to writing a code patch. The deep learning-based vector representation of issues in DeepSoft provides an opportunity to revisit some existing well-known problems such as detecting bug duplicate reports or bug localization. In addition, it also opens new research opportunities related to the highly challenging problem of automatically generating code patches for resolving an issue,  and code and API recommendation in general.  Code generation refers to the production of a sequence of code tokens given a context vector. Given the recent successes in NLP \cite{manning2016computational}
(machine translation, question answering, and dialog systems) and vision  \cite{lecun2015deep} (image/video captioning/story telling and more recently, visual question answering), it is expected that the technologies can be leveraged. In \cite{sutskever2014sequence} for example, LSTM is used to generate target language sentences given a context vector that encodes the source language sentence.

\subsection{Release planning}

Since DeepSoft models the progression of issues (including bugs and new feature request), it can be applied to: (i) make recommendations about the new functionalities that should be implemented next; or (ii) predicting how the software product quality (e.g. in terms of defects) will evolve over time. The former leads to a new, data-driven approach to the well-known next release problem where existing approaches focus only on using evolutionary optimization techniques. The latter introduces a new approach to the defection prediction problem which no longer requires manual feature engineering.


\subsection{Predicting effort and risks}
Effort and risk prediction can be made at each level in DeepSoft. Using the deep learning-based vector representation of an issue, we can build more accurate models for estimating the effort of resolving it or predicting whether its resolution will be delayed. Similar risk prediction can be made at the release and project level. For example, DeepSoft can be applied to predict whether new releases will be delivered in time and meeting the target.


\subsection{Developer modeling}

Developers' involvement dictates the path of software evolution, thus representing developers is critical. In DeepSoft, a developer can be represented as a sequence of issues he or she has involved with, and can therefore be modelled as an irregular-time LSTM. This capability suggests many applications such as developer recommendation in bug triaging, or developer productivity and team capability estimate.

%% file: fse-vision-2016.bbl
\begin{thebibliography}{10}

\bibitem{Andreessen2011}
M.~Andreessen.
\newblock Why software is eating the world.
\newblock {\em The Wall Street Journal}, August 2011.

\bibitem{Choetkiertikul2015}
M.~Choetkiertikul, H.~K. Dam, T.~Tran, and A.~Ghose.
\newblock {Predicting delays in software projects using networked
  classification}.
\newblock In {\em Proceedings of the 30th IEEE/ACM International Conference on
  Automated Software Engineering (ASE)}, pages 353 -- 364, 2015.

\bibitem{hochreiter1997long}
S.~Hochreiter and J.~Schmidhuber.
\newblock Long short-term memory.
\newblock {\em Neural computation}, 9(8):1735--1780, 1997.

\bibitem{Le2014}
Q.~V. Le and T.~Mikolov.
\newblock Distributed representations of sentences and documents.
\newblock In {\em Proceedings of the 31th International Conference on Machine
  Learning{ICML}}, volume~32 of {\em {JMLR} Proceedings}, pages 1188--1196.
  JMLR.org, 2014.

\bibitem{lecun2015deep}
Y.~LeCun, Y.~Bengio, and G.~Hinton.
\newblock Deep learning.
\newblock {\em Nature}, 521(7553):436--444, 2015.

\bibitem{manning2016computational}
C.~D. Manning.
\newblock Computational linguistics and deep learning.
\newblock {\em Computational Linguistics}, 2016.

\bibitem{Menzies:2010}
T.~Menzies, Z.~Milton, B.~Turhan, B.~Cukic, Y.~Jiang, and A.~Bener.
\newblock Defect prediction from static code features: Current results,
  limitations, new approaches.
\newblock {\em Automated Software Engg.}, 17(4):375--407, Dec. 2010.

\bibitem{MikolovSCCD13}
T.~Mikolov, I.~Sutskever, K.~Chen, G.~S. Corrado, and J.~Dean.
\newblock Distributed representations of words and phrases and their
  compositionality.
\newblock In {\em Advances in Neural Information Processing Systems 26: 27th
  Annual Conference on Neural Information Processing Systems 2013.}, pages
  3111--3119, 2013.

\bibitem{mou2016convolutional}
L.~Mou, G.~Li, L.~Zhang, T.~Wang, and Z.~Jin.
\newblock Convolutional neural networks over tree structures for programming
  language processing.
\newblock In {\em Proceedings of the Thirtieth AAAI Conference on Artificial
  Intelligence}, 2016.

\bibitem{Nguyen:2016}
T.~D. Nguyen, A.~T. Nguyen, and T.~N. Nguyen.
\newblock Mapping api elements for code migration with vector representations.
\newblock In {\em Proceedings of the 38th International Conference on Software
  Engineering Companion}, ICSE '16, pages 756--758, New York, NY, USA, 2016.
  ACM.

\bibitem{DeepCare2016}
T.~Pham, T.~Tran, D.~Phung, and S.~Venkatesh.
\newblock Deepcare: A deep dynamic memory model for predictive medicine.
\newblock In {\em Advances in Knowledge Discovery and Data Mining. Proceeding
  of the 20th Pacific Asia Conference on Knowledge Discovery and Data Mining.
  Lecture Notes in Computer Science}, volume 9652 of {\em PAKDD'16}, pages
  30--41, New York, NY, USA, 2016. Springer.

\bibitem{Schork2015}
N.~J. Schork.
\newblock Personalized medicine: Time for one-person trials.
\newblock {\em Nature}, 520(7549):609--611, April 2015.

\bibitem{sutskever2014sequence}
I.~Sutskever, O.~Vinyals, and Q.~V. Le.
\newblock Sequence to sequence learning with neural networks.
\newblock In {\em Advances in Neural Information Processing Systems}, pages
  3104--3112, 2014.

\bibitem{Zimmermann:2009:CDP}
T.~Zimmermann, N.~Nagappan, H.~Gall, E.~Giger, and B.~Murphy.
\newblock Cross-project defect prediction: A large scale experiment on data vs.
  domain vs. process.
\newblock In {\em Proceedings of the the 7th Joint Meeting of the European
  Software Engineering Conference and the ACM SIGSOFT Symposium on The
  Foundations of Software Engineering}, ESEC/FSE '09, pages 91--100, New York,
  NY, USA, 2009. ACM.

\end{thebibliography}
